\documentclass[showpacs,showkeys,preprintnumbers,twocolumn,amsmath,amssymb,groupedaddress,superscriptaddress]{revtex4-1}

\usepackage{graphicx}

\usepackage{dcolumn}

\usepackage{bm}

\begin{document}

%\draft

\title{The absolute mass of neutrino and the first 
unique forbidden $\beta$-decay of ${^{187}}{\rm Re}$}

\author{Rastislav Dvornick\'y}
\affiliation{Department of Nuclear 
Physics and Biophysics, Comenius University, Mlynsk\'a dolina F1, 
SK--842 15
Bratislava, Slovakia}
\affiliation{Bogoliubov Laboratory of Theoretical Physics, JINR Dubna,
141980 Dubna, Moscow region, Russia}
\author{Kazuo Muto}
\affiliation{Department of Physics, Tokyo Institute of Technology,
Tokyo 152-8551, Japan }
\author{Fedor \v Simkovic}
\affiliation{Department of Nuclear 
Physics and Biophysics, Comenius University,
Mlynsk\'a dolina F1, SK--842 15
Bratislava, Slovakia}
\affiliation{Bogoliubov Laboratory of Theoretical Physics, JINR Dubna,
141980 Dubna, Moscow region, Russia}
\author{Amand Faessler}
\affiliation{Institute f\"{u}r Theoretische Physik der Universit\"{a}t
T\"{u}bingen, D-72076 T\"{u}bingen, Germany}

\date{\today}

\begin{abstract}
The planned rhenium $\beta$-decay experiment, called the 
``Microcalorimeter Arrays for a Rhenium Experiment'' 
(MARE), might probe the absolute mass scale of neutrinos with the same
sensitivity as the Karlsruhe tritium neutrino mass (KATRIN) experiment, 
which will take commissioning data in 2011 and will proceed for
5 years. We present the energy distribution of emitted electrons
for the first unique forbidden $\beta$-decay of ${^{187}}{\rm Re}$.
It is found that the $p$-wave emission of electron dominates
over the $s$-wave. 
By assuming mixing of three neutrinos the Kurie function for the rhenium 
$\beta$-decay is derived.  It is shown that 
the Kurie plot near the endpoint is within a good accuracy linear
in the limit of massless neutrinos like the Kurie plot of the 
superallowed $\beta$-decay of ${}^{3}{\rm H}$.
\end{abstract}



\date{\today}

\nopagebreak[4]

\maketitle 

\section{Introduction}

The recent atmospheric, solar, reactor and accelerator neutrino 
experiments have convinced us that neutrinos are massive particles.
However, the problem of absolute values of their masses is still
waiting for a solution.
Neutrino oscillations  depend on the differences of neutrino masses,
not on their absolute values.
Apparently three kinds of neutrino 
experiments have a chance to determine the light neutrino masses: 
i) cosmological measurements,
ii) the tritium and rhenium single $\beta$-decay experiments,
iii) neutrinoless double $\beta$-decay experiments.

The measurement of the electron spectrum in $\beta$-decays provides
a robust direct determination of the values of neutrino masses. 
In practice, the most sensitive experiments use tritium $\beta$-decay,
because it is a super-allowed transition with a low $Q$-value. 
The effect of neutrino masses $m_k$ ($k$=1,2,3) can be observed 
near the end point of the electron spectrum, where $Q-T \sim m_k$.
$T$ is the electron kinetic energy.
A low $Q$-value is important,
because the relative number of events occurring in an interval
of energy ${\mit\Delta} T$ near the endpoint is proportional
to $({\mit\Delta}T/Q)^3$.

The current best upper bound on the effective neutrino mass 
$m_\beta$ given by,
\begin{equation}
m_\beta = \sqrt{\sum_{k=1}^3 |U_{ek}|^2 m^2_k},
\end{equation}
has been obtained in the Mainz and Troitsk experiments: 
$m_\beta < 2.2~{\rm eV}$ \cite{otten}.
$U_{ek}$ is the element of neutrino mixing matrix.
In the near future, the Karslruhe tritium neutrino mass  
(KATRIN) experiment will reach a sensitivity of about 
$0.2~{\rm eV}$ \cite{katrin}. In this experiment the $\beta$-decay 
of tritium will be investigated with a spectrometer
taking advantage of magnetic adiabatic collimation 
combined with an electrostatic filter.

Calorimetric measurements of the $\beta$-decay of rhenium  where 
all electron energy released in the decay is recorded,
appear complementary to those carried out with spectrometers. 
The unique first forbidden $\beta$-decay,
\begin{equation}
{^{187}}{\rm Re} \rightarrow {^{187}}{\rm Os} + e^- + {\overline{\nu}}_e,
\end{equation} 
is particularly promising due to its low transition energy of
$\sim 2.47~{\rm keV}$ and the large isotopic abundance of
${}^{187}{\rm Re}$ ($62.8$\%),
which allows the use of absorbers made with natural rhenium.
Measurements of the spectra 
of ${}^{187}{\rm Re}$ have been reported by the Genova and
the Milano/Como groups (MIBETA and MANU experiments).
The achieved  sensitivity of  
$m_\beta < 15~{\rm eV}$ was limited by statistics \cite{otten}.
The success of rhenium experiments has encouraged 
the micro-calorimeter community to proceed with
a competitive precision search for a neutrino mass.
The ambitious project, called the ``Microcalorimeter Arrays
for a Rhenium Experiment'' (MARE), is planned in two steps,
MARE I and MARE II. MARE I might reach a statistical
sensitivity of 4 eV after 3 years of data taking \cite{mare}.  
MARE II is to challenge the KATRIN goal of 0.2 eV \cite{mare}.

The aim of this paper is to derive the form of the endpoint 
spectrum of emitted electrons for the $\beta$-decay of ${}^{187}{\rm Re}$,
which is needed to extract the effective neutrino mass $m_\beta$
or to place a limit on this quantity  from future MARE I and II experiments.

\section{First unique forbidden $\beta$-decay of ${}^{187}{\rm Re}$}

The ground-state spin-parity is $5/2^{+}$ for ${}^{187}{\rm Re}$
and $1/2^{-}$ for the daughter nucleus ${}^{187}{\rm Os}$,
and the decay is associated with ${\mit\Delta}J^{\pi} = 2^{-}$
(${\mit\Delta}L = 1$, ${\mit\Delta}S = 1$) of the nucleus, {\it i.e.},
classified as first unique forbidden $\beta$-decay. The emitted
electron and neutrino are expected to be, respectively, in 
$p_{3/2}$ and $s_{1/2}$ states (see Appendix A) or vice versa. 
The emission of higher partial waves is strongly 
suppressed due to a small energy release in this nuclear
transition.

\vspace{1.0cm}
\begin{figure}[htb!]
%    \epsfxsize=0.45\textwidth
%    \epsffile{srcpaper_fig1.eps}
\includegraphics[width=.47\textwidth,angle=0]{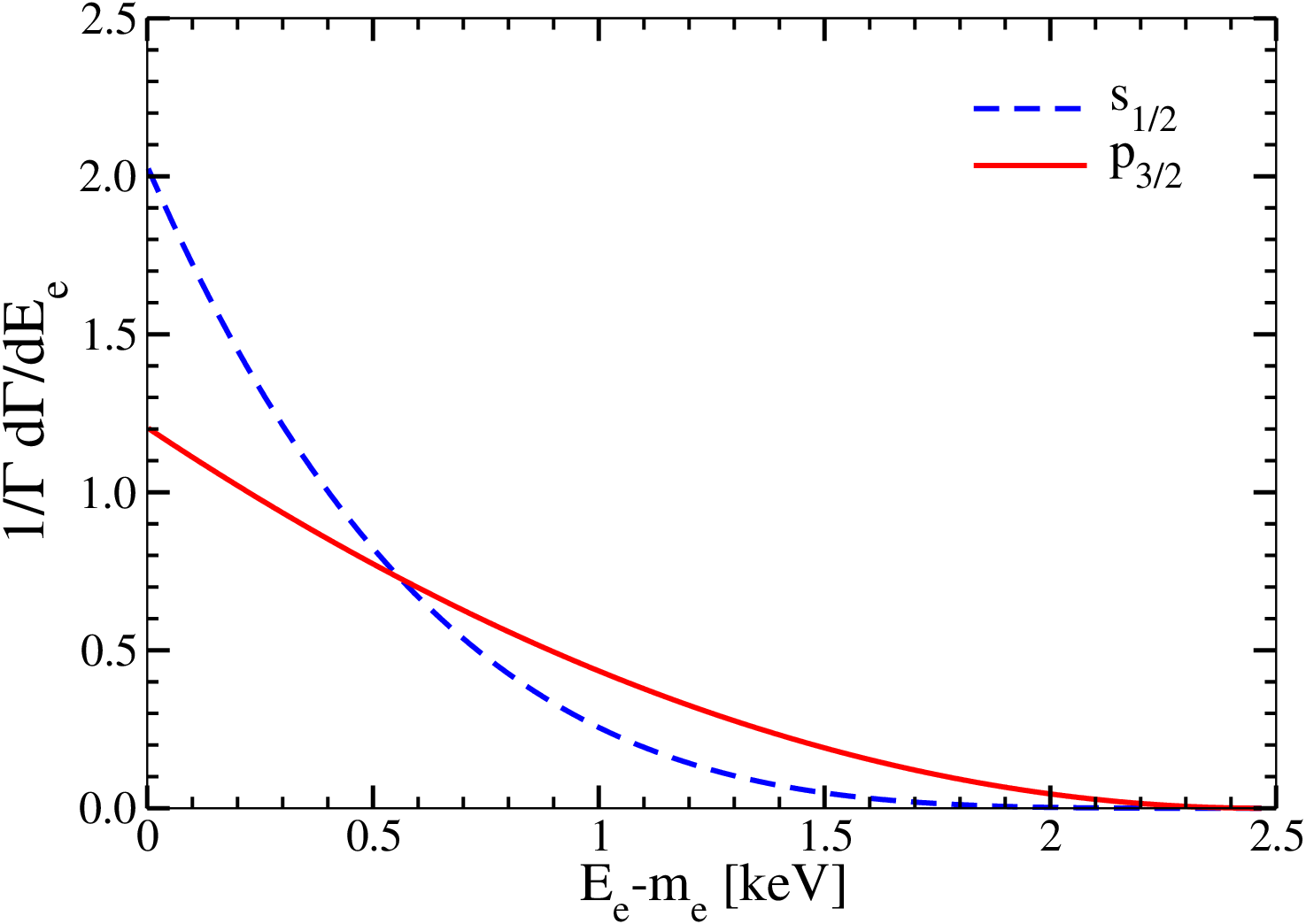}
\vspace{0.5cm}
  \caption{(Color online) The single electron differential decay rate
normalized to the particular decay rate 
(${\mit\Gamma}^{s_{1/2}}$ or ${\mit\Gamma}^{p_{3/2}}$) for emission
of $s_{1/2}$ and $p_{3/2}$ electrons 
vs. electron energy $E_e$ for $\beta$-decay 
of ${}^{187}{\rm Re}$.} 
\label{fig:1}
\end{figure}

The differential decay rate is a sum of two contributions associated
with emission of the $s_{1/2}$ and the $p_{3/2}$ state electrons. 
By considering the finite nuclear size effect the theoretical spectral
shape of the $\beta$-decay of ${^{187}{\rm Re}}$ is
\begin{eqnarray}
\frac{d {\mit\Gamma}}{dE_e} &=& \frac{d {\mit\Gamma}^{p_{3/2}}}{dE_e} 
+ \frac{d {\mit\Gamma}^{s_{1/2}}}{dE_e} \nonumber\\
&=& \sum_{k=1}^3 |U_{ek}|^2~ 
\frac{G_F^2 V_{ud}^2}{2 \pi^3}~
  B ~R^2 ~p_e~ E_e ~(E_0 - E_e)
\nonumber\\ 
&\times&  \frac{1}{3}
  \left[ F_{1}(Z,E_e)p_e^2 + F_{0}(Z,E_e)((E_0 - E_e)^2-m^2_{k})
\right]
\nonumber\\
&& ~~\times \sqrt{(E_0 - E_e)^2-m^2_{k}}~~ 
\theta(E_0-E_e-m_k) 
  \label{eq.sp}
\end{eqnarray}
with
\begin{eqnarray}
B &=& \frac{g_A^2}{6 R^2} \label{nme}
|\langle {1/2}^{-} \Vert \sum_n
\tau^+_n 
\{ \bm{\sigma}_{n}^{~} \otimes \bm{r}_{n}^{~} \}_{2}
\Vert {5/2}^{+}\rangle|^2.{~~~~}
\end{eqnarray}
$G_F$ is the Fermi constant and $V_{ud}$ is the element of the
Cabbibo-Kobayashi-Maskawa (CKM) matrix. $p_e$, $E_e$ and $E_0$
are the momentum, energy, and maximal endpoint energy (in
the case of zero neutrino mass) of the electron, respectively.
$F_{0}(Z,E)$ and $F_{1}(Z,E)$ in Eq. (\ref{eq.sp})
are relativistic Fermi functions and $\theta(x)$ is a  
theta (step) function. $g_A$ denotes an axial-vector 
coupling constant. $\bm{r}_{n}^{~}$ is a coordinate of the ${\rm n}$-th nucleon
and $R$ is a nuclear radius.
The value of nuclear matrix element $B$ in Eq. (\ref{nme}) 
can be determined from the measured half-life
of the $\beta$-decay of ${}^{187}{\rm Re}$.

%\vspace{2cm}
\begin{figure}[htb!]
%    \epsfxsize=0.45\textwidth
%    \epsffile{srcpaper_fig1.eps}
\includegraphics[width=.45\textwidth,angle=0]{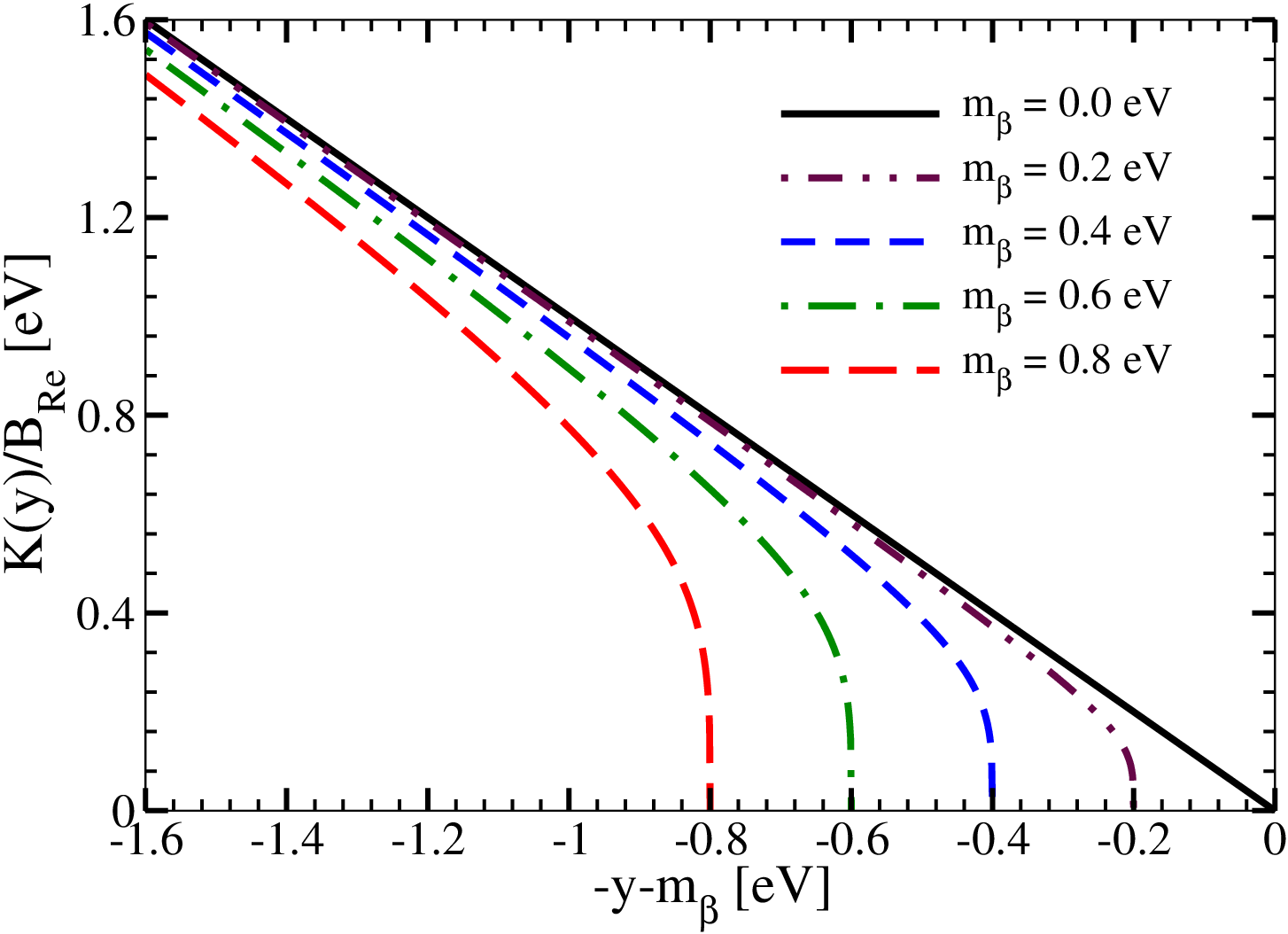}
%\vspace{0.3cm}
  \caption{(Color online) 
Endpoints of the Kurie plot of the rhenium $\beta$-decay
for various values of the effective neutrino mass:
$m_\beta = 0,~ 0.2,~ 0.4,~ 0.6$ and $0.8$ eV.} 
\label{fig:2}
\end{figure}

The $\beta$-decay rate of rhenium is a sum of 
decay rates for emission of $p_{3/2}$ and $s_{1/2}$ electrons
[see Eq. (\ref{eq.sp})]. 
In Fig.~\ref{fig:1} we show the single electron differential 
decay rate normalized to the particular decay rate as a 
function of electron energy $E_e$. The two possibilities offer 
a different energy distribution of outgoing electrons.
Close to the end point, there is a more flat distribution for
$s_{1/2}$ electrons due to the dependence on squared 
neutrino momentum $p^2_\nu = (E_0 - E_e)^2-m^2_{k}$
[see Eq. (\ref{eq.sp})].
Because of this factor the two particular decay rates
depend on the neutrino mass in a different way.

Experimentally, it was found that $p_{3/2}$-state electrons are predominantly
emitted in the $\beta$-decay of $^{187}{\rm Re}$ \cite{andrea}. By performing
numerical analysis of partial decay rates associated with emission 
of the $p_{3/2}$ and the $s_{1/2}$ electrons (terms with
$F_{1}(Z,E_e)$  and $F_{0}(Z,E_e)$ in Eq. (\ref{eq.sp}), respectively) 
we conclude that about $10^{4}$ times more $p_{3/2}$-state electrons are emitted 
when compared with emission of $s_{1/2}$-state electrons. The reasons for
it are as follows: i) $F_{1}(Z,E_e) \gg F_{0}(Z,E_e)$ for $E_e-m_e < Q$ 
(see Appendix B),
ii) the maximal momentum of electron ($\sim 49.3$ keV) is much larger
than the maximal momentum of the neutrino ($\sim 2.5$ keV). Henceforth,
we shall neglect a small contribution to the differential decay rate given by 
an emission of  the $s_{1/2}$-state electrons.

For a normal hierarchy (NH) of neutrino masses  with $m_3 > m_2 > m_1$ 
the Kurie function of the $\beta$-decay of $^{187}{\rm Re}$ is given by
\begin{eqnarray}
K (y) &=&
\sqrt{\frac{d\Gamma/dE_e}{p_e E_e F_0(Z,E_e)}}
\nonumber\\
&=& {\cal B}_{Re} \sqrt{ y + m_1}
\Bigl[ |U_{e1}|^2 \sqrt{y ( y + 2 m_1 )}  \Bigr.\nonumber\\
&&  + |U_{e2}|^2 \sqrt{ ( y + m_1 - m_2) (y + m_1 + m_2)} \times
\nonumber\\
&& ~~~~~~~~\theta(y + m_1 - m_2) \nonumber\\
&& + |U_{e3}|^2 \sqrt{ (y + m_1 - m_3) ( y + m_1 + m_3 )}\times
\nonumber\\
&& ~~~~~~~~\Bigl.  \theta(y + m_1 - m_3) \Bigr] ^{1/2},
\label{kpre1}
\end{eqnarray}
with
\begin{eqnarray}
{\cal B}_{Re} = \frac{G_F V_{ud} \sqrt{B} }{\sqrt{2 \pi^3}}
\sqrt{\frac{R^2~p_e^2}{3}
\frac{F_{1}(Z,E_e)}{F_{0}(Z,E_e)}} 
\label{BRe}
\end{eqnarray}
and $y = (E_0 - E_e - m_1) \ge 0$. The ratio 
$\left(p^2_e ~F_{1}(Z,E_e)\right)/F_{0}(Z,E_e)$ depends only weakly on the
electron momentum in the case of the $\beta$-decay of rhenium.
With a good accuracy the factor ${\cal B}_{Re}$ can be considered 
to be a constant. 

The current upper limit on neutrino mass from tritium and
rhenium $\beta$-decay experiments holds in the degenerate 
neutrino mass region ($m_1 \simeq m_2 \simeq m_3 \simeq m_0$
with $m_0 = \sum_{i=1}^3 m_i/3$) for which $m_\beta \simeq m_0$.  
So far the future rhenium $\beta$-decay experiments will not see 
any effect due to a smallness of neutrino masses, it is 
possible to approximate $m_k \ll Q - T$ (k=1,2,3) and obtain
\begin{eqnarray}
K(y) \simeq  {\cal B}_{Re} \left( (y + m_\beta) 
\sqrt{ y ( y + 2 m_\beta ) } \right)^{1/2},
\end{eqnarray}
where $y = (E_0 - E_e - m_\beta )$. In Fig.~\ref{fig:2} 
we show the Kurie plot for
$\beta$-decay of ${}^{187}{\rm Re}$ versus $y$ near the endpoint. We see that
the Kurie plot is linear near the endpoint  for $m_\beta = 0$. However,
the linearity of the Kurie plot is lost if $m_\beta$ is not equal to zero. 

We note that there is also a possibility of bound-state decay of ${}^{187}{\rm Re}$,
where an electron is placed into a bound state above occupied electron shells 
of the final atom. In Ref. \cite{Wil83} it was shown that the ratio of decay rates of  
bound-state and continuum $\beta$-decays of ${}^{187}{\rm Re}$ is less than $1\%$. 

\section{Conclusions}

For the first unique forbidden $\beta$-decay of ${}^{187}{\rm Re}$ 
to ground state of ${}^{187}{\rm Os}$ the theoretical spectral shape
is presented. The decay rate of the process is a sum of particular
decay rates associated with emission of $s_{1/2}$ and $p_{3/2}$ 
electrons, which depend in a different way on the neutrino mass.
The p-wave emission is dominant over the s-wave.  So, the Kurie 
function, defined by Eq. (\ref{BRe}), is almost linear in the
endpoint region. An observed deviation from the linearity indicates 
effects of the finite neutrino mass.

The analysis of the the Kurie function of the first forbidden
$\beta$-decay of ${}^{187}{\rm Re}$ show that close to the endpoint 
it coincides up to a factor to the Kurie function 
of superallowed $\beta$-decay of tritium \cite{tritium}.
These findings are important for the planned MARE experiment,
which will be sensitive to neutrino mass in the sub eV region.

\section*{Acknowledgments}

This work was supported in part by the DFG project 436 SLK 17/298 
and by the VEGA Grant agency  under the contract No.~1/0249/03.

\appendix
\section{Relativistic electron wave functions}

\begin{figure}[t!]
%    \epsfxsize=0.45\textwidth
%    \epsffile{srcpaper_fig1.eps}
\includegraphics[width=.47\textwidth,angle=0]{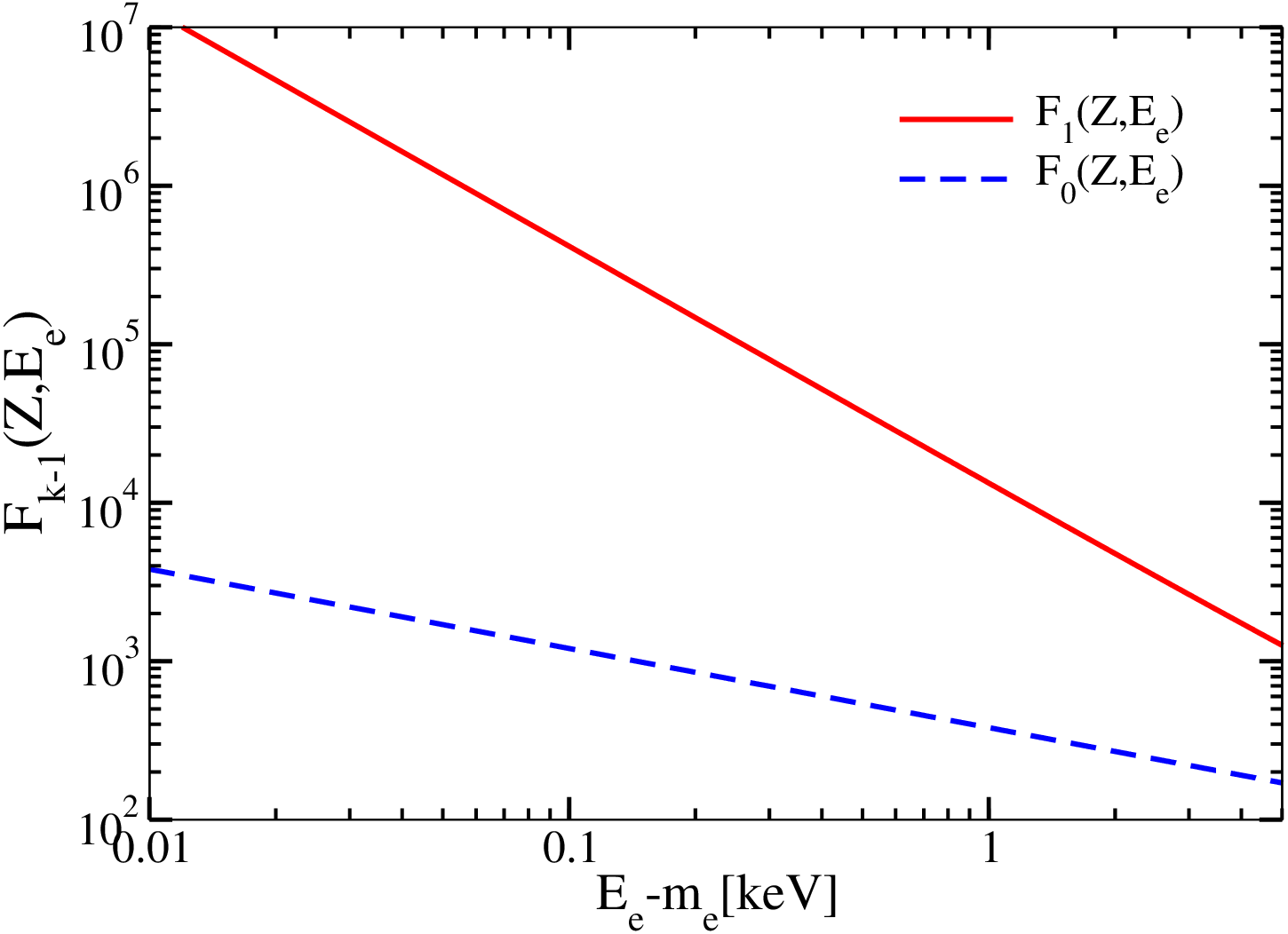}
%\vspace{0.3cm}
  \caption{(Color online) 
Relativistic Fermi functions $F_0(Z,E_e)$
and $F_1(Z,E_e)$ vs. electron energy $E_e$ 
for the $\beta$-decay of ${}^{187}{\rm Re}$.}
\label{fig:3}
\end{figure}

We adopt the relativistic electron wave function in a
uniform charge distribution in a nucleus, which is 
expanded in terms of spherical waves:
\begin{eqnarray}
\Psi(E_e,\bm{r}) &=& \Psi^{(s_{1/2})}(E_e,\bm{r})  \nonumber\\ 
&& + \Psi^{(p_{1/2})}(E_e,\bm{r}) + \Psi^{(p_{3/2})}(E_e,\bm{r}) + \cdots
\end{eqnarray} 
where of particular interest are $s_{1/2}$ and $p_{3/2}$
states \cite{doi}:
\begin{eqnarray}
\Psi^{(s_{1/2})}(E_e,\bm{r}) &=& 
\left(\begin{array}{l}
{\tilde g}_{-1 }(r)\chi_s \\
{\tilde f}_{+1 }(r) (\mathbf{\sigma\cdot\hat{p}}) \chi_s
\end{array}\right) \nonumber\\
\Psi^{(p_{3/2})}(E_e,\bm{r}) &=& 
i \left(\begin{array}{l}
{\tilde g}_{-2 }(r)[3 (\mathbf{\hat{r}\cdot\hat{p}})
-(\mathbf{\sigma\cdot\hat{r}})(\mathbf{\sigma\cdot\hat{p}})]
\chi_s \\
{\tilde f}_{+2}(r)
[(\mathbf{\hat{r}\cdot\hat{p}})
(\mathbf{\sigma\cdot\hat{p}})
- (\mathbf{\sigma\cdot\hat{r}})] 
\chi_s
\end{array}\right).\nonumber\\
\end{eqnarray}
Here, $\bm{r}$ is a position vector, $r=|\bm{r}|$, $\hat{\bm{r}}= \bm{r}/r$  
and $\hat{\bm{p}} = {\bm{p}}_e/p_e$. 
By keeping the lowest power in expansion of $r$ 
the radial wave functions take the form:
\begin{equation}
\left(\begin{array}{l}
{\tilde g}_{-1 }(r) \\
{\tilde f}_{+1 }(r)
\end{array}\right) = {\tilde A}_{\mp 1},
~~~
\left(\begin{array}{l}
{\tilde g}_{-2}(r) \\
{\tilde f}_{+2}(r)
\end{array}\right) = {\tilde A}_{\mp 2}  (p r/3).
\end{equation}
In approximation up to $(\alpha Z)^2$ terms we have
\begin{equation}
{\tilde A}_{\pm k} \simeq \sqrt{(E_e\mp m_e)/(2E_e)}
\sqrt{F_{k-1}(Z,E_e)}.
\end{equation}

\section{Relativistic Fermi functions}

Relativistic Fermi function $F_{k-1}(Z,E_e)$
takes into account the distortion
of electron wave function due to electromagnetic
interaction of the emitted electron with the atomic nucleus.
It takes the form \cite{doi}:
\begin{eqnarray}
F_{k-1}(Z,E_e)
 & = & \left( \frac{{\mit\Gamma} (2k+1)}
          {{\mit\Gamma}(k) {\mit\Gamma}(1+2\gamma_k)}  \right)^2 
(2 p_e R )^{2(\gamma_k -k)} \nonumber \\
 & \times &  |{\mit\Gamma} (\gamma_k +iz)|^2 e^{\pi z} 
\end{eqnarray}

where $k=1,2,3, \cdots$ and
\begin{eqnarray}
\gamma_k &=& \sqrt{k^2-(\alpha Z)^2} \nonumber \\
z &=&  \alpha Z \frac{E_e}{p_e}.
\end{eqnarray}

Following Ref. \cite{doi91} it can be written as
\begin{eqnarray}
F_{k-1}(Z,E_e) &=& C_{k-1} ~d_{k-1}(E_e) ~ 
\left( \frac{m_e}{p_e} \right)^{2k-1}
\left( \frac{E_e}{m_e} \right)^{2\gamma_k -1}.
\nonumber\\
\end{eqnarray}
The constant $C_{k-1}$ is given by
\begin{eqnarray}
C_{k-1} &=& 2 \pi \alpha Z 
(2 \alpha Z m_e R )^{2(\gamma_k-k)}
\left[ \frac{2k (2k-1)!!}{{\mit\Gamma}(1+2\gamma_k)}  \right]^2.
\nonumber\\
\end{eqnarray}
The function $d_{k-1}$ takes the form
\begin{eqnarray}
d_{k-1}(E_e) &=& \frac{1}{2\pi} ~z^{1-2\gamma_k}~ e^{\pi z}
~ |{\mit\Gamma} (\gamma_k +iz)|^2.
\end{eqnarray}
We note that 
\begin{eqnarray}
\lim_{p_e \rightarrow 0} d_{k-1}(E_e) = 1. 
\end{eqnarray}

The Fermi functions $F_0(Z,E_e)$ and $F_1(Z,E_e)$ are 
related to the emission of $s_{1/2}$ and $p_{3/2}$ electrons,
respectively. In Fig. \ref{fig:3}, they are plotted as functions
of the kinetic energy of electrons emitted in $\beta$-decay of $^{187}Re$.
We note that values of $F_1(Z,E_e)$ are significantly larger 
than those of $F_0(Z,E_e)$.

A quantity of interest is the ratio
\begin{eqnarray}
\frac{p^2_e F_1(Z,E_e)}{F_0(Z,E_e)} &=&
\frac{C_1}{C_0} m_e^2~ \left( \frac{E_e}{m_e} \right)^{2(\gamma_2-\gamma_1)}
\nonumber\\
&\simeq& \frac{C_1}{C_0} m_e^2~ 
\left[ 1 + 2 \frac{(E_e-m_e)}{m_e} \right].
\end{eqnarray}
As the $Q$-value of rhenium $\beta$-decay is only $2.47$ keV the above ratio
can be to a good accuracy considered to be a constant with 
the value $0.11~ m_e^2$.

\end{document}